\documentclass{appolb}
\usepackage{epsfig}
\usepackage{subfigure}
\usepackage{amsmath}

\begin{document}
\title{Investigating the origin of the long range pseudo rapidity correlation in 2d di-hadron measurements from STAR%
\thanks{Presented at Strange Quark Matter conference, September, 2011, Krakow, Poland.}%
}
\author{L. C. De Silva
\address{The Department of Physics, University of Houston,
617 SR Building 1,
Houston, Texas 77204-5005}
}
\maketitle
\begin{abstract}
Angular di-hadron correlations reveal novel structures in central Au+Au collisions at $\sqrt{S_{NN}}$ = 200 GeV. One of them, known as the ridge, is elongated in pseudo rapidity and peaks on the near side ($\Delta\phi$ $\approx$ 0). Investigating the origin of the ridge structure helps to understand the hot dense matter that is created in ultra relativistic heavy ion collisions. Results showing the  $\langle$$p_{T}$$\rangle$ dependence of the ridge structure are presented. Evidence for possible jet and non-jet contributions to the ridge structure will be discussed.
\end{abstract}

\section{Introduction}
Recent triggered di-hadron correlation studies by STAR report a ridge structure in two dimensions ($\Delta\eta$,$\Delta\phi$)\cite{jana}. Based on their studies, the near side was assumed to consist of two independent structures, a jet-like structure and the ridge. A complementary approach, presented here, is carried out by using all possible charged particle pairs. The approach does not require a trigger particle but with appropriate kinematic cuts reproduces qualitatively similar correlation structure to that of triggered analysis. In this approach, two empirical 2d fit functional models are used to extract ridge properties. Possible model dependent quantitative evidence for jet and non-jet phenomena contribution to the near side structure can be extracted from the fits.

\section{Data and analysis}
The data used in this analysis were collected during run 4, year 2004, from the STAR detector at Relativistic Heavy Ion Collider (RHIC) Brookhaven National Lab (BNL), Long Island, New York. 32M Au+Au collisions at $\sqrt{S_{NN}}$ = 200 GeV  were analyzed. Charged tracks reconstructed using the Time Projection Chamber (TPC) with 2$\pi$ azimuthal coverage and $|\eta|$ $\leq$ 1 in pseudo rapidity were selected. An earlier centrality evolution analysis\cite{daugherity} was based on particles in the full transverse momentum range above 0.15 GeV/c. For the study presented here the lower threshold of the transverse momentum was raised for both particles. The selected event vertexes are within $\pm$ 25cm of the detector center and primary tracks are selected to come from within a distance of closet approach of 3cm to the event vertex.  

The correlation function for this analysis is a construct of PearsonÕs correlation coefficient. Mathematically it can be derived as\cite{daugherity}:

\begin{align}
          \frac{\Delta\rho}{\sqrt{\rho_{ref}}} &= \frac{\rho_{sib}\--\rho_{mix}}{\sqrt{\rho_{ref}}}
\end{align}

$\rho_{sib}$ is the normalized charged particle pair density within a single event, which carries both correlated and uncorrelated pairs. The uncorrelated background is subtracted by mixed event pairs. The resulting uncorrelated density is denoted as $\rho_{mix}$. The difference is the number of correlated pairs which is normalized by the square root of mixed pair density to yield the \emph{correlated particle pairs per final state charged particle} as the equivalent correlation measure to the PearsonÕs correlation coefficient. In order to avoid possible artifacts due to pseudo rapidity acceptance, the correlation measure is carried out in sub bins of the z-vertex. The event mixing artifacts were eliminated by mixing events of a particular centrality within a multiplicity window of 50 charged tracks. To correct for pair loss a pair cut is implemented on both mixed and same event pairs. The effects due to pileup events were removed by one of the standard pileup eliminating procedures in STAR\cite{duncan}. Finally the conversion electron positron pair background is reduced by using a 1.5$\sigma$ dE/dX cut on electrons in momentum ranges, 0.2 \textless $p_{T}$ \textless 0.45 GeV/c and 0.7 \textless $p_{T}$ \textless 0.8 GeV/c.

\begin{figure}[ht]
\centering
\includegraphics[width=0.5\textwidth]{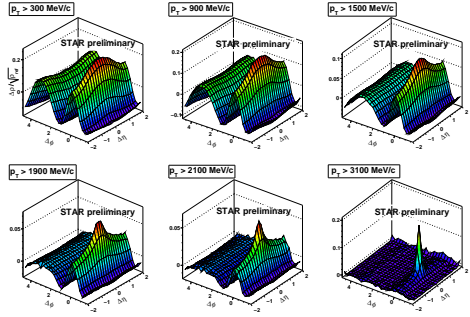}
\caption{Evolution of the raw normalized correlation structure for selected $\langle$$p_{T}$$\rangle$ bins at 0-1\% centrality bin. The lower momentum threshold of both particles have been increased in generating the spectrum of correlation plots.}
\label{figure:one}
\end{figure}

\section{Fit functions}
The initial empirical fit function carries three mathematical model components that lead to seven free fit parameters.

F = $c_{0}$ + $c_{1}$cos($\Delta\phi$) + $c_{2}$exp(-0.5*(($\Delta\phi$/$c_{3}$)$^2$ + ($\Delta\eta$/$c_{4}$)$^2$) + $c_{5}$exp(-0.5*(($\Delta\phi$/$c_{6}$)$^2$ + ($\Delta\eta$/$c_{6}$)$^2$))

The first component of the fit is an offset parameter followed by a cos($\Delta\phi$) term to extract the away side structure as seen in Fig.~\ref{figure:one}. The assumption is that this structure corresponds to correlations due to a recoiling jet\cite{porter}. The two remaining terms are introduced based on $\langle$$p_{T}$$\rangle$ evolution of data (see Fig.~\ref{figure:one}). The asymmetric 2d Gaussian attempts to address the long range correlation properties excluding the peak structure around (0,0) that appears as  $\langle$$p_{T}$$\rangle$ increases. This structure is described via the symmetric 2d Gaussian component of the fit.

\begin{figure}[h]
\centering
\includegraphics[width=0.7\textwidth]{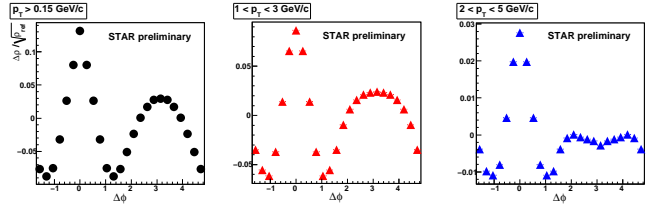}
\caption{Correlations at high $p_{T}$ very central data show evidence for the existence of higher order harmonics.}
\label{figure:two}
\end{figure}

Modifications to the model were introduced after indications of higher order harmonics were first predicted \cite{xin,alver} and then observed (see Fig.~\ref{figure:two}). 

F = $c_{0}$ + $c_{1}$cos($\Delta\phi$) + $c_{2}$cos(2$\Delta\phi$) + $c_{3}$cos(3$\Delta\phi$) + $c_{4}$cos(4$\Delta\phi$) + $c_{5}$cos(5$\Delta\phi$) + $c_{6}$exp(-0.5*(($\Delta\phi$/$c_{7}$)$^2$ + ($\Delta\eta$/$c_{8}$)$^2$))

The resulting fit carries higher order harmonics up to 5th order and an asymmetric 2d Gaussian. It is important to note that in such a Fourier expansion, the first order term serves as a momentum conserving term (correlation due to recoil jet). In essence, the new model help in further constraining the knowledge gathered from the initial model study.

\section{Results}
The focus of comparing the initial model study to the study which includes fit components for higher harmonics based on initial energy density fluctuations is to determine whether any information can be extracted from the asymmetric 2d Gaussian fit to the away side structure. 
\subsection{Initial model study}
\begin{figure}[h]
        \centering
        \subfigure[Amplitude]
       {
        \includegraphics[width=0.31\textwidth]{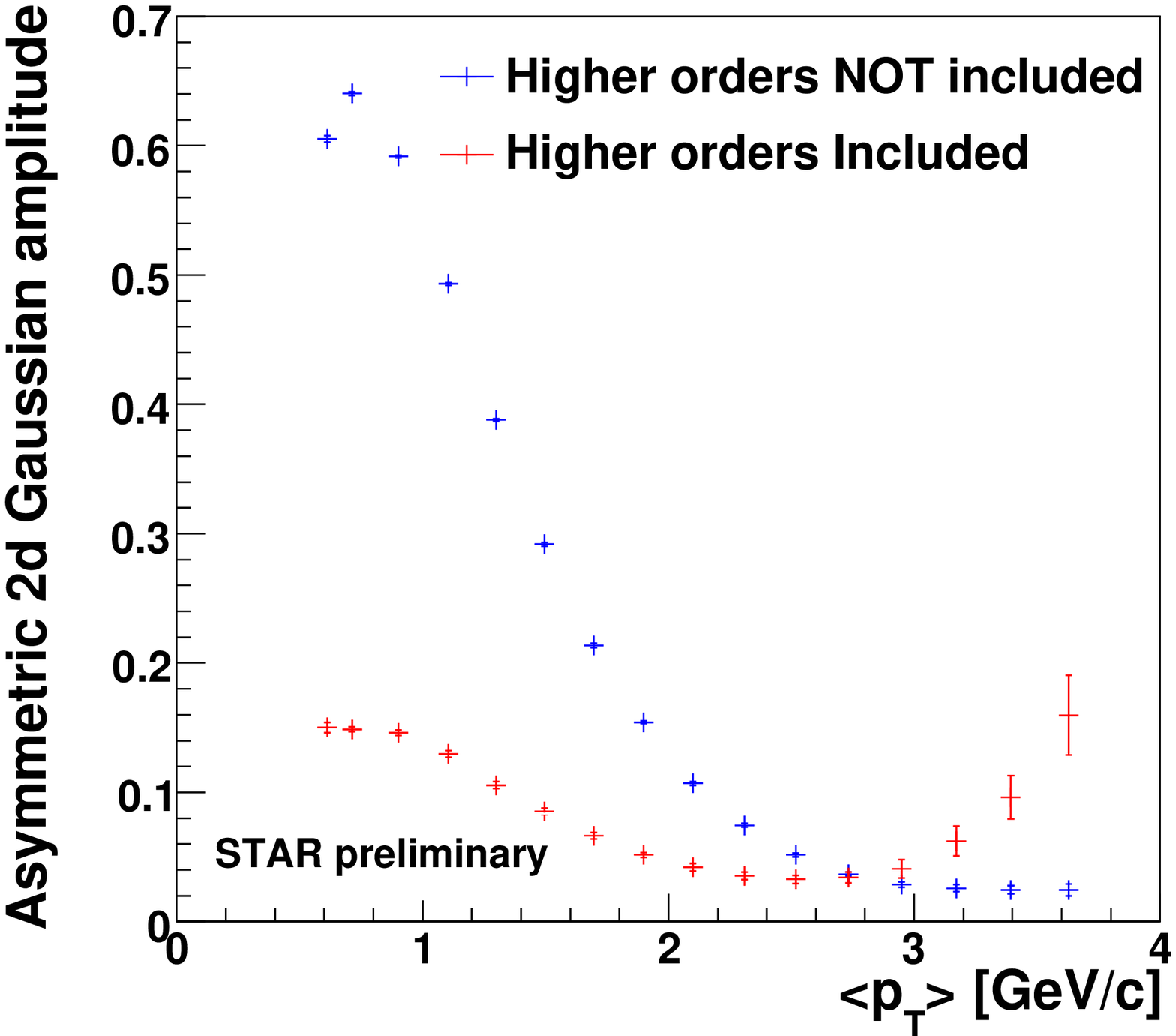}
        \label{figure:3a}
       }
       \subfigure[$\Delta\eta$ width]
      {
       \includegraphics[width=0.31\textwidth]{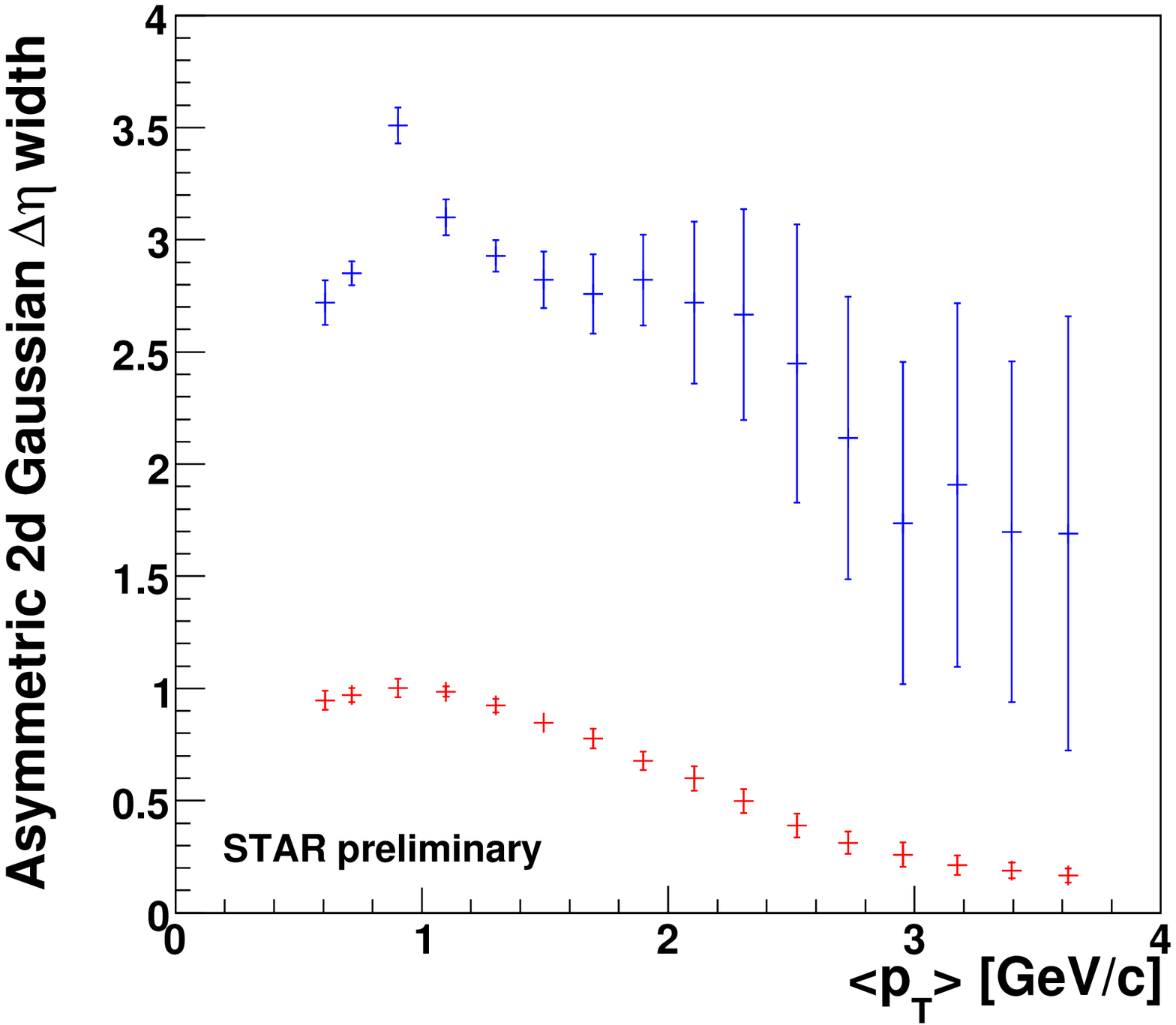}
       \label{figure:3b}
      }
      \\
      \subfigure[$\Delta\phi$ width]
     {
      \includegraphics[width=0.31\textwidth]{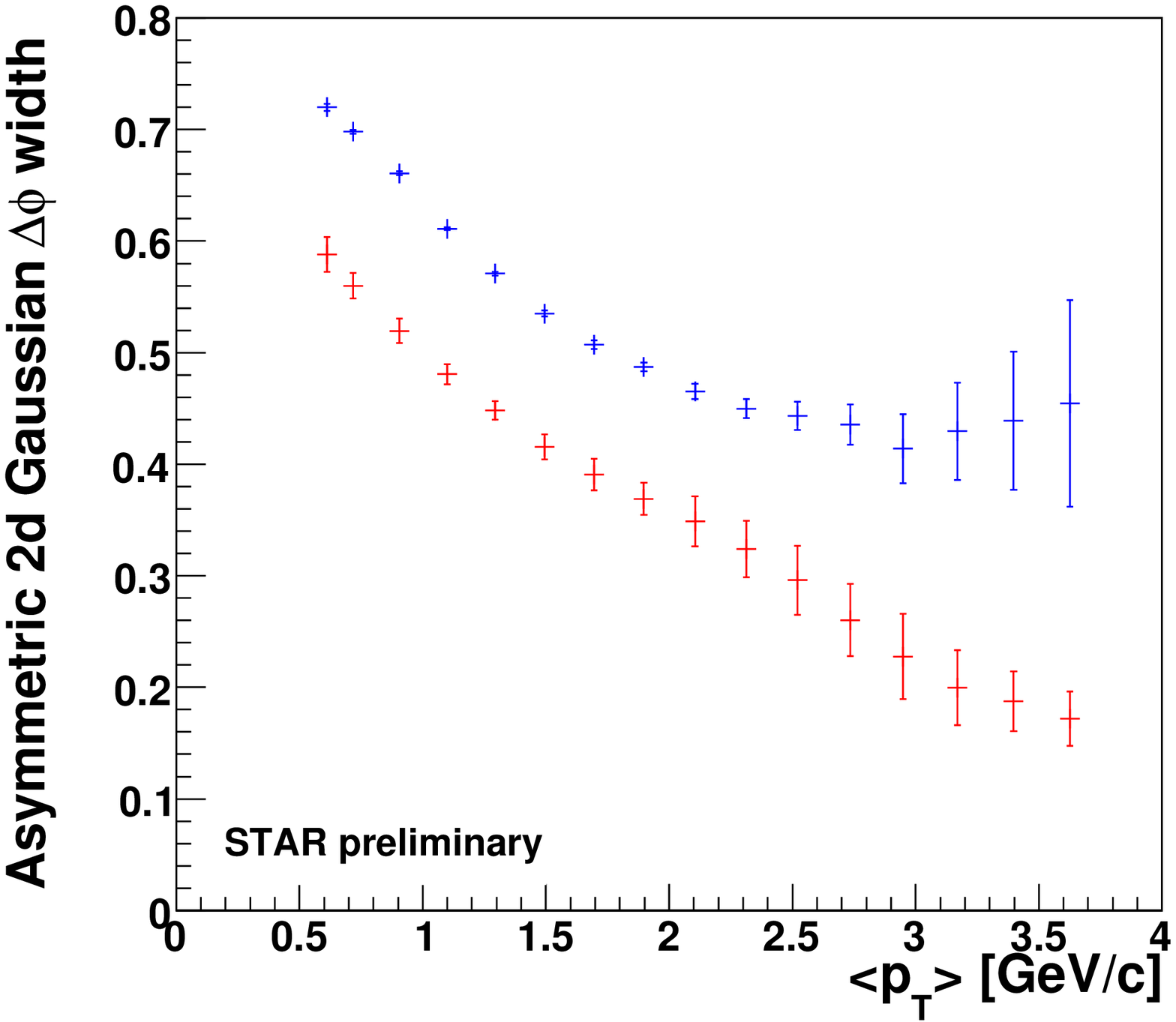}
      \label{figure.3c}
      }
      \subfigure[Symmetric width]
      {
      \includegraphics[width=0.30\textwidth]{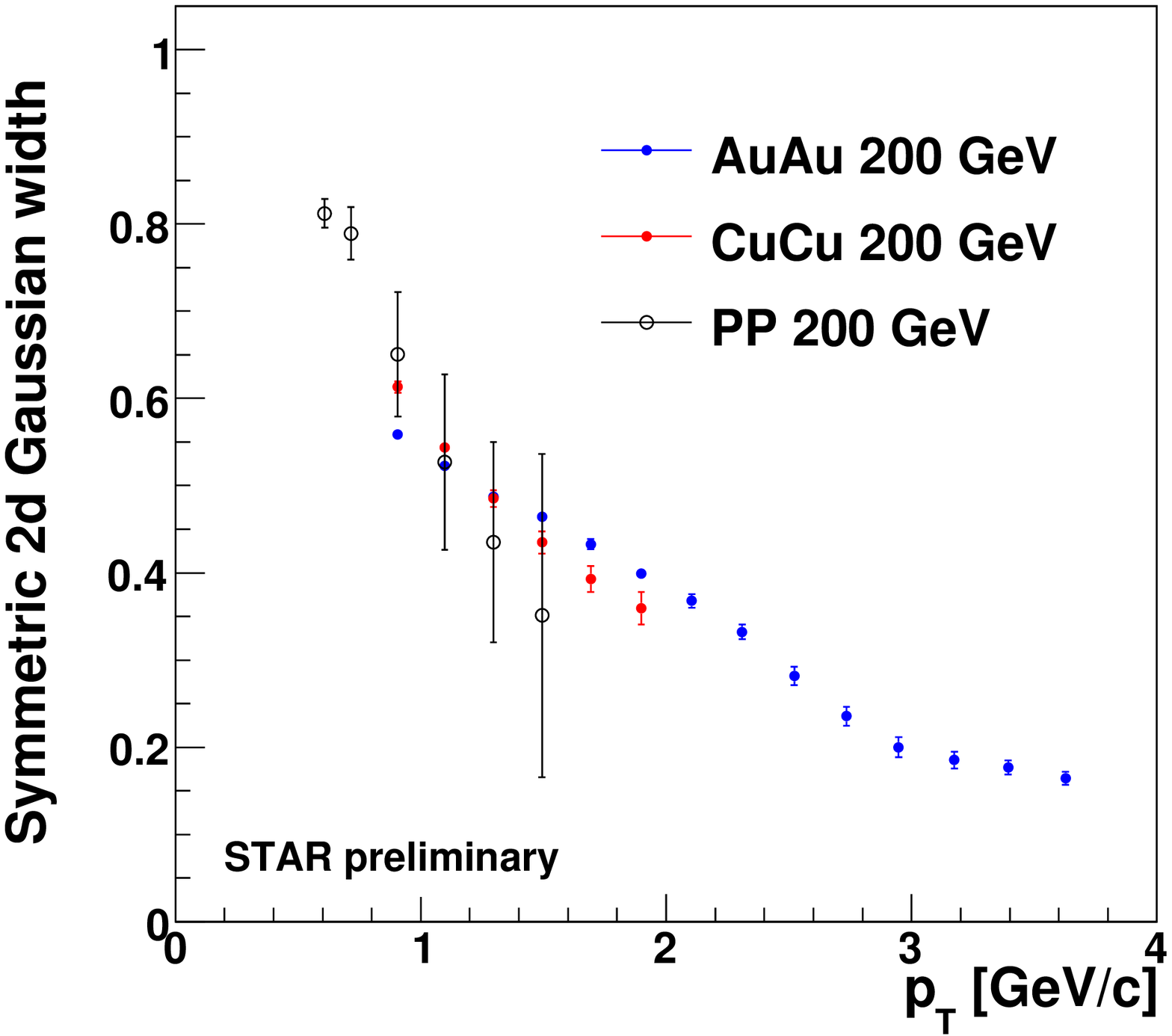}
      \label{figure:3d}
      }
      \caption{Asymmetric and Symmetric 2d Gaussian parameter evolution as a function of $\langle$$p_{T}$$\rangle$.}
      \label{figure:three}
\end{figure}

Resulting asymmetric 2d Gaussian parameters are reported in Fig.~\ref{figure:three}. The long correlation strength approaches zero at higher $\langle$$p_{T}$$\rangle$ as indicated by the asymmetric 2d Gaussian amplitude and volume parameters. The $\Delta\eta$ and $\Delta\phi$ width evolution clearly indicate the asymmetry shown in data. Also the $\Delta\eta$ width suggests that within the STAR acceptance the ÒridgeÓ is flat whereas the $\phi$ width shows a very smooth narrowing. An important observation is that the $\langle$$p_{T}$$\rangle$ of the ridge ($\langle$$p_{T}$$\rangle$ = 0.72 GeV/c) is greater than that of the inclusive spectrum ($\langle$$p_{T}$$\rangle$ $\approx$ 0.42 GeV/c \cite{spectra}). The hardness of the ridge spectrum is an indication of possible Jet contributions to the ridge. 

On the other hand, the symmetric 2d Gaussian component reveals important information about an unmodified jet emerging at higher momenta (Fig.~\ref{figure:three}). The width is comparable to the jet width in PP collisions. In order to further constrain the above conclusion, amplitude and yield comparisons need to be made.

\begin{figure}[h]
      \centering
      \subfigure[$\frac{v_{3}}{v_{2}}$ ratio]
      {
       \includegraphics[width=0.33\textwidth]{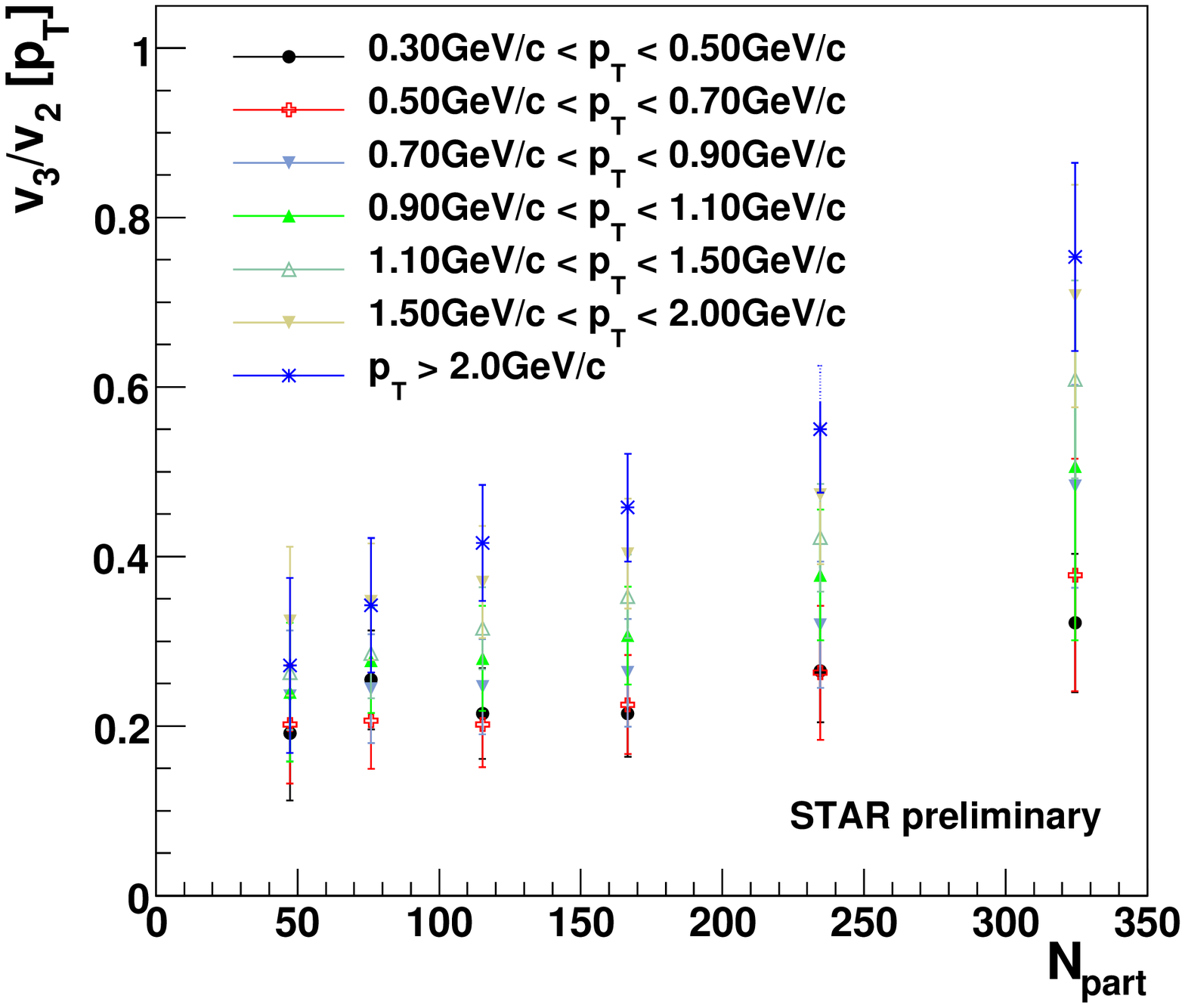}
       \label{figure:3a}
      }
      \subfigure[$v_{n}\hspace{0.1 cm} scaling\hspace{0.1 cm}at\hspace{0.1 cm}0\--10\%$]
      {
       \includegraphics[width=0.30\textwidth]{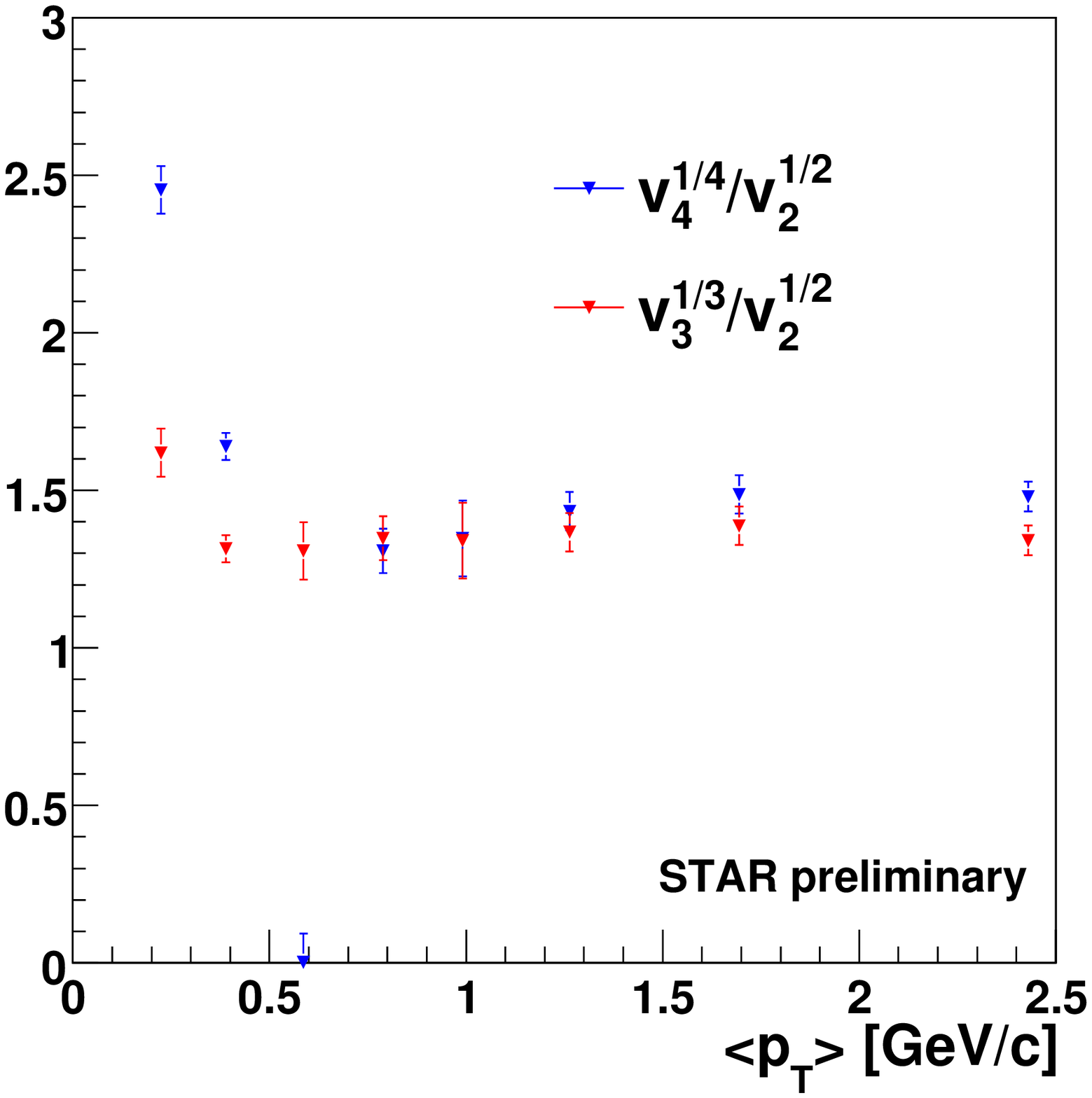}
       \label{figure:3a}
      }
      \caption{Comparison to predicted higher harmonic ratios\cite{alver} and scaling relations\cite{xin}}
      \label{figure:four}
\end{figure}

\subsection{Higher harmonics study}
Recent advancements in heavy ion collision models, predict significant contributions from higher order harmonics to the observed correlation spectra due to initial energy density fluctuations\cite{xin,alver}. STAR data also support the predicted observation in high $p_{T}$, very central AuAu 200 GeV data(Fig.~\ref{figure:two}). The $\frac{v_{3}}{v_{2}}$ ratio predictions were tested and reasonable agreement to theory is evident (see Fig.~\ref{figure:four}a). However it is to be noted at low $p_{T}$, deviations from the trends are observed for $N_{part}$ \textless 150. One should note, thought, that the data is extracted fitting 2d di-hadron correlations, whereas, theory uses single particle spectra and event plane calculations.  

A predicted hydro scaling relation\cite{clement}  using extracted higher order Fourier coefficients is studied, and reasonable agreement to theory has been observed (Fig.~\ref{figure:four}b).In summary, given the different approaches used by theory predictions tested in this proceedings and experiment, the data trends seem support the hydro based theory predictions to a reasonable extent. 

The remaining structure on the near side, after subtracting the Fourier components still shows a $\Delta\eta - \Delta\phi$ asymmetry (see Fig.~\ref{figure:three}, red data markers). This $\langle$$p_{T}$$\rangle$ evolution indicates that this $\Delta\eta$ elongation is visible up to 2.7 GeV/c and is strongest at 0.9 GeV/c. The structure thus suggests possible jet modification at low $\langle$$p_{T}$$\rangle$ which evolves to an unmodified jet at high $p_{T}$. The amplitude and volume of the asymmetric structure scales lineary with Glauber scaling as a function of centrality, which hints at a jet origin. The rise of amplitude at high $\langle$$p_{T}$$\rangle$ is expected due to an increase in per charge particle pair yield in jets. Further studies will be carried out using the comparison to p+p collisions.

\section{Summary and discussion}

\begin{figure}[ht]
\centering
\includegraphics[width=0.30\textwidth]{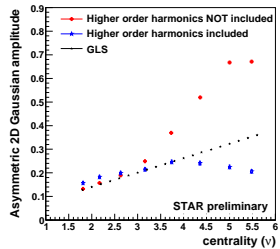}
\caption{Comparison of near side asymmetric 2D Gaussian amplitude to Glauber linear scaling as a function of centrality.}
\label{figure:five}
\end{figure}

Un-triggered 2d di-hadron correlation studies reproduce qualitatively similar results to that of a triggered analysis. The observed near side correlation is modeled via an empirical fit function which extracts short and long range structure properties. The $\langle$$p_{T}$$\rangle$ evolution of the extracted parameters suggests a possible correlation between jets and the observed long range correlation. Further constraint to data has been imposed modifying the empirical model to incorporate higher order Fourier model components cos(n$\Delta\phi$), n = 1\--- 5, based on theoretical predictions and evidence in the data. The extracted higher order component strengths show reasonable agreement to predicted hydrodynamical trends. The remainder on the near side still reveals an asymmetric 2d Gaussian suggestive of possible modified jet phenomena. A Glauber linear scaling was successfully applied to the un-triggered centrality evolution (see Fig.~\ref{figure:five}) which indicates a jet origin for the observed near side Gaussian. 

\begin{thebibliography}{5}   

\bibitem{jana} B. I. Abelev et al. (STAR Collaboration), Phys. Rev. {\bf C 80}, 064912 (2009)

\bibitem{daugherity} M Daugherity et al. (STAR Collaboration), J. Phys. {\bf G 35}, 104090 (2008)

\bibitem{porter} J. Porter and T. Trainor, J. Phys.: Conf. Ser. {\bf 27}, 98 (2005) 

\bibitem{xin} Xin Niang Wang et al., Phys. Rev. Lett. {\bf 106}, 162301(2011)

\bibitem{alver} B. Alver et al., Phys. Rev. {\bf C 81}, 054905 (2010)

\bibitem{clement} C. Gombeaud et al., Phys. Rev. {\bf C 81}, 014901 (2010)

\bibitem{duncan} Nuclear Phys. Lab. Annual Report, University of Washington (2009) p. 58

\bibitem{spectra} J. Adams et al. STAR Collaboration

\end{thebibliography}
\end{document}